\documentclass[letterpaper]{article} 
\usepackage{aaai23}  
\usepackage{times}  
\usepackage{helvet}  
\usepackage{courier}  
\usepackage[hyphens]{url}  
\usepackage{graphicx} 
\usepackage{natbib}  
\usepackage{caption} 
\usepackage{algorithm}
\usepackage{algorithmic}

\usepackage{cite}
\usepackage{verbatim}
\usepackage{amsmath}
\usepackage{amsmath}
\usepackage{booktabs}
\usepackage{newfloat}
\usepackage{listings}

\usepackage{amsmath}
\usepackage{multirow}
\usepackage{multicol}
\usepackage{color}

\usepackage{subfigure}
\usepackage{times}
\usepackage{epsfig}
\usepackage{amssymb}
\usepackage{xcolor}
\usepackage{bm}
\usepackage{array}
\usepackage{caption}
\DeclareCaptionStyle{ruled}{labelfont=normalfont,labelsep=colon,strut=off} 
\floatstyle{ruled}
\newfloat{listing}{tb}{lst}{}
\floatname{listing}{Listing}
\title{Fast Online Hashing with Multi-Label Projection}

\title{Fast Online Hashing with Multi-Label Projection}
\author {
    Wenzhe Jia\textsuperscript{\rm 1,2},
    Yuan Cao\thanks{Corresponding author}\textsuperscript{\rm 1,2},
    Junwei Liu\textsuperscript{\rm 1},
    Jie Gui \textsuperscript{\rm 3,4}
}
\affiliations {
    \textsuperscript{\rm 1} Ocean University of China, China\\
    \textsuperscript{\rm 2} State Key Laboratory of Integrated Services Networks (Xidian University), China\\
    \textsuperscript{\rm 3} Southeast University, China\\
    \textsuperscript{\rm 4} Purple Mountain Laboratories, China\\
    jiawenzhe@stu.ouc.edu.cn, cy8661@ouc.edu.cn, liujunwei@stu.ouc.edu.cn, guijie@seu.edu.cn
}

\usepackage{bibentry}

\begin{document}

\maketitle

\begin{abstract}
Hashing has been widely researched to solve the large-scale approximate nearest neighbor search problem owing to its time and storage superiority. In recent years, a number of online hashing methods have emerged, which can update the hash functions to adapt to the new stream data and realize dynamic  retrieval. However, existing online hashing methods are required to update the whole database with the latest hash functions when a query arrives, which leads to low retrieval efficiency with the continuous increase of the stream data. On the other hand, these methods ignore the supervision relationship among the examples, especially in the multi-label case. In this paper, we propose a novel Fast Online Hashing (FOH) method which only updates the binary codes of a small part of the database. To be specific, we first build a query pool in which the nearest neighbors of each central point are recorded. When a new query arrives, only the binary codes of the corresponding potential neighbors are updated. In addition, we create a similarity matrix which takes the multi-label supervision information into account and bring in the multi-label projection loss to further preserve the similarity among the multi-label data. The experimental results on two common benchmarks show that the proposed FOH can achieve dramatic superiority on query time up to 6.28 seconds less than state-of-the-art baselines with competitive retrieval accuracy. \end{abstract}

\section{Introduction}
With the increasing amount of data available on the Internet, Approximate Nearest neighbor (ANN) search \cite{wang2017survey} has achieved a widespread success in many applications, e.g. computer vision and cross-modal retrieval problems. Hashing-based methods \cite{wang2012semi,liu2016sequential,lu2017deep} have attracted extensive attention for ANN search due to their advantages in terms of data storage and computational efficiency. Hashing aims at mapping high-dimensional features into compact binary codes, while preserving similarities between the original space and the binary space.

Most of the existing popular hashing methods are based on batch-learning strategy \cite{he2019k,cao_sdish}, which hinders their ability to adapt to changes as a dataset grows and diversifies, because the computational cost may become intractable and infeasible. Hence, online hashing methods have emerged, which demonstrate good performance-complexity trade-offs by updating hash functions from streaming data \cite{cakir2017mihash}. Online hashing focuses on updating hash functions and hash tables constantly on the basis of continual stream data with low cost \cite{cakir2017online,lu2019efficient,wang2020label}.

Online hashing can be generally divided into unsupervised hashing \cite{leng2015online,chen2017frosh} and supervised hashing \cite{lin2020hadamard,fang2021label}. Unsupervised online hashing is roughly based on the idea of ``sketching" \cite{clarkson2009numerical}. The sketch is a smaller feature matrix that preserves the main features of the database. By realizing matrix decomposition, the hash functions can be updated dynamically and efficiently. Supervised online hashing is mainly based on two kinds of supervision information: similarity matrix and label. The former produces a similarity matrix based on the supervision information. Then, the loss function is constructed by approximating the inner product of the paired data and corresponding similarities. The latter generates a codebook and aims at assigning each codeword a unique label, so that data with the same label will have approximate hash codes.

Although existing supervised online hashing methods update hash functions efficiently, the hash table is updated too frequent to obtain high search efficiency. Specifically speaking, since the hash functions are updated constantly, the whole hash table needs to be updated based on the latest hash functions, when a new query arrives. Otherwise, the query is embedded by the latest hash functions, but the hash codes of the database are based on previous hash functions, which is not symmetric and leads to low accuracy without doubt. However, updating the whole hash table is too time consuming with the increasing database, which is one of the core problems in online hashing.

\begin{figure}[t]
	\centering
	\includegraphics[scale=0.4]{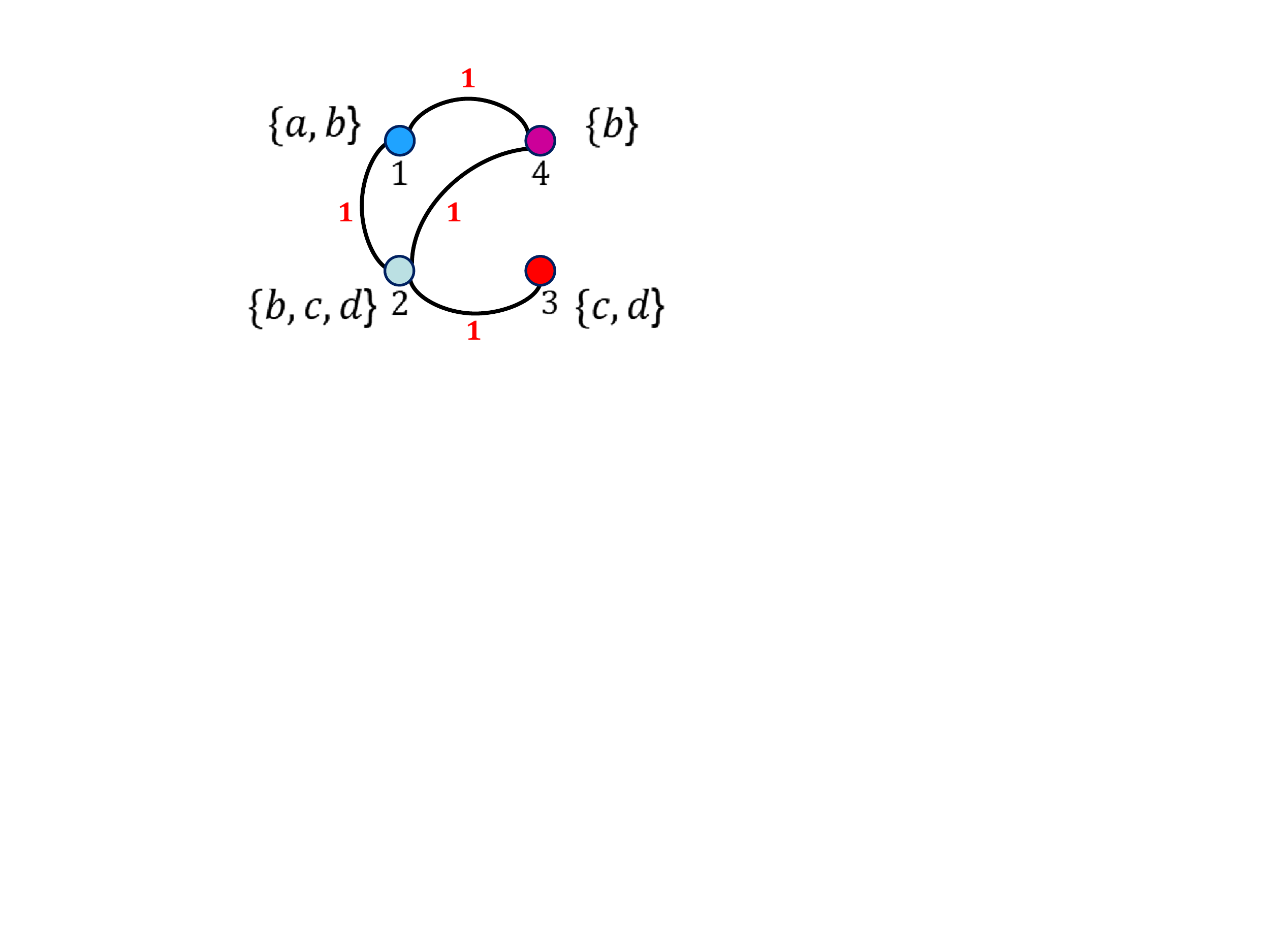}
	\vspace{-2.2in}
	\caption{An example to show the similarities of the existing supervised hashing methods in the multi-label case. \{a,b,c,d\} denotes the label information, the red numbers denotes the similarities.}
	\label{example}
	\vspace{-0.25in}
\end{figure}

On the other hand, most of the existing supervised hashing methods contribute to construct a codebook and assign each codeword a unique label. This strategy ignores the similarity relationship among the examples, especially in the multi-label case. For example, Fig. \ref{example} shows the label information of four points and the similarities among them. Most of the existing methods consider two examples the same (similarity equals 1) if they share at least one common label, otherwise, similarity equals 0 (no edge exists between two points in Fig. \ref{example}). Obviously, the similarity between the 3rd point and the 2nd point should be higher than that between the 4th point and the 2nd point. However, the existing methods consider these two kinds of cases the same, which is not reasonable. Besides, most of these methods just take one kind of similarity criterion into consideration, i.e., similarity matrix or label, which neglects the construction of different angels of loss functions.

In this paper, we propose a novel Fast Online Hashing (FOH) method based on multi-label projection. In order not to update the hash codes of all the database when a new query arrives, we build a query pool by randomly sampling a few central points from the database. Besides, we present a neighbor-preserving algorithm to record the nearest neighbors of the central points. The central points in the query pool and corresponding neighbors are updated based on the stream data according to the reservoir sampling strategy. In this way, when a new query arrives, a few nearest central points of the query in the query pool are returned and the corresponding potential nearest neighbors are recorded. Only the recorded data points are required to be embedded into hash codes based on the latest hash functions. Finally, Hamming distances are computed between the query hash code and the hash codes of the recorded potential nearest neighbors to obtain the retrieval results.

In addition, we make full use of the label supervision information of the data to generate the hash codes. As for the multi-label data, we present a construction algorithm to create the similarity matrix in consideration of the multi-label information of the data. Furthermore, both the similarity matrix and the label information are applied to the construction of the final loss function, which conduces to higher retrieval accuracy. To summarize, the main contributions of the proposed FOH approach are as follows.

\vspace{-0.05in}
\begin{itemize}
    \item A query pool is introduced to preserve the potential nearest neighbors of the query, which makes the query time reduced. The neighbor-preserving algorithm and the reservoir sampling strategy guarantee the true neighbors of the query not to be omitted.
    \vspace{-0.05in}
    \item As for multi-label supervision information, a novel similarity matrix is created to further preserve similarities among the examples. Besides, the label projection loss is brought in the final loss function.
    \vspace{-0.05in}
    \item Experimental results show dramatic query time superiority in comparison to state-of-the-art supervised online hashing methods with competitive retrieval accuracy.
\end{itemize}

\vspace{-0.1in}
\section{Related Work}
One of the representative unsupervised online hashing methods is Online Sketch Hashing (SketchHash) \cite{leng2015online}. SketchHash uses a sketch matrix to preserve the main features of the data. The hash learning task is transformed into calculating several maximum eigenvalues and eigenvectors of the sketch matrix. In order to speed up the matrix decomposition, Faster Online Sketch Hashing (FROSH) \cite{chen2017frosh} downexamples the random Hadamard transform to speed up the training of the sketch matrix.

Supervised online hashing learns hash functions based on the label information, which can narrow the semantic gap. Online Kernel Hashing (OKH) \cite{huang2013online} is the first attempt to update hash functions in a paired-input fashion. With an online passive-aggressive strategy, important information about the stream data is maintained. AdaptiveHash \cite{cakir2015adaptive} defines a hinge-loss function and optimises the model dramatically based on SGD. Mutual Information Hashing (MIH) \cite{cakir2017mihash} utilizes mutual information as the objective function and updates the hash tables based on it. Balanced Similarity for online Discrete Hashing \cite{lin2019towards} investigates the correlation between the existing data and the new data. BSODH sets two balancing factors to solve the ``imbalance problem" caused by the asymmetric graphs and optimises them by means of discretization. Hadamard Matrix Guided Online Hashing (HMOH) \cite{lin2020hadamard} considers the Hadamard matrix as a more discriminative codebook. By assigning each column of the Hadamard matrix a unique label as target, the hash functions are updated.

\begin{figure*}[t]
	\centering
	\includegraphics[scale=0.27]{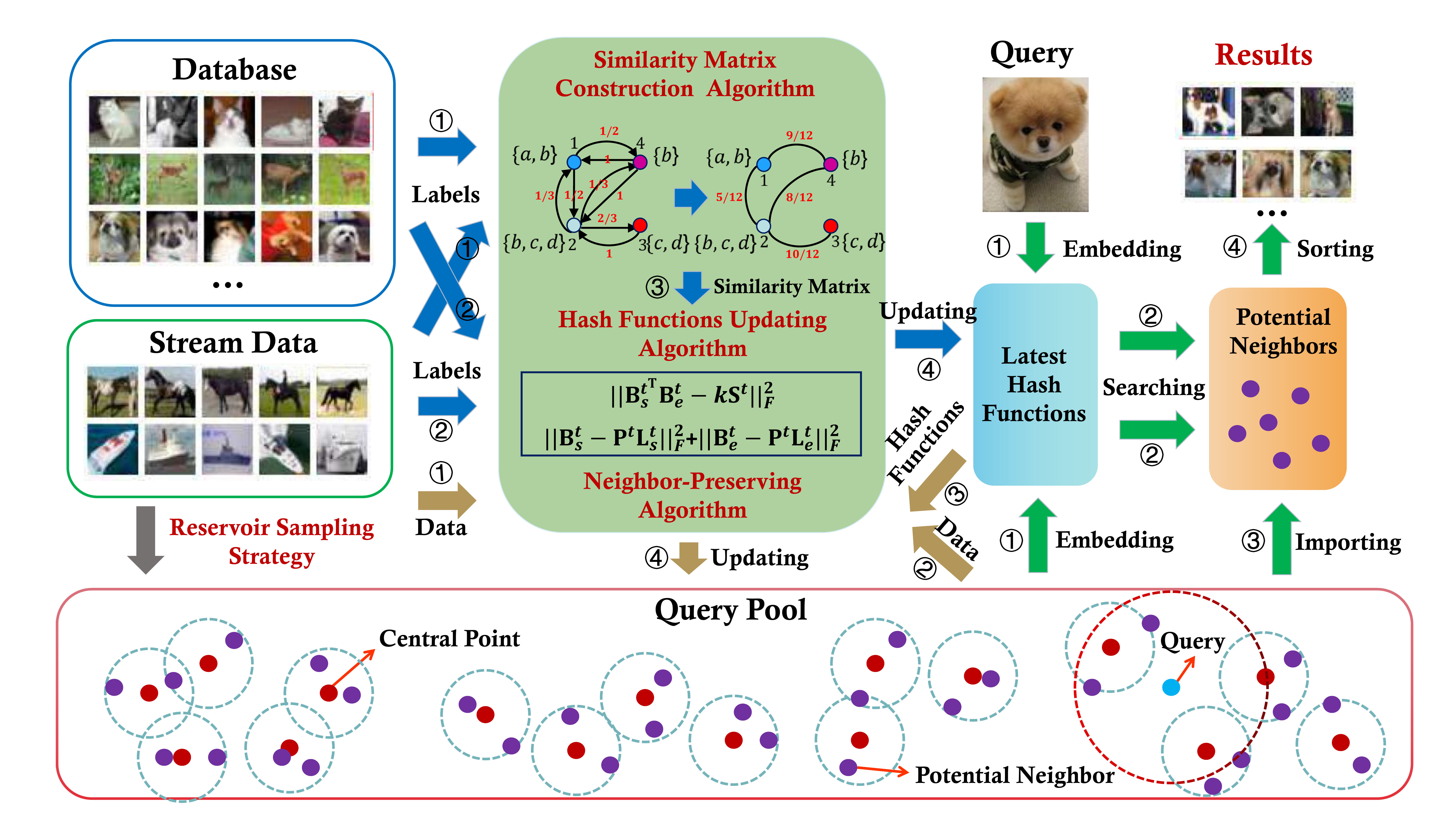}
	\vspace{-0.05in}
	\caption{The whole framework of the proposed Fast Online Hashing model. The process consists of four parts. The blue arrows denote the hash functions updating process. The yellow arrows denote the updating of the nearest neighbors of the central points in the query pool. The gray arrow denotes the reservoir sampling strategy to update the central points. The green arrows denote the online query process.}
	\label{framework}
	\vspace{-0.2in}
\end{figure*}

With the rapid growth of multi-modal data, multi-modal online hashing has appeared \cite{xie2017dynamic,yi2021efficient}. These methods concentrate on searching for semantically related examples from one modality (e.g. image) with queries from another modality (e.g. text) in the stream data condition. The first proposed cross-modal online hashing (OCMH) \cite{xie2016online} learns shared latent matrices and variable matrices for each modality, enabling efficient update of the hash codes. Flexible Online Multi-modal Hashing (FOMH) \cite{lu2019flexible} learns the modal combination weights adaptively based on the online streaming multi-modal data. The learnt self-weighted multi-modal fusion hash codes allow flexible fusion of heterogeneous modalities even if some of them are lost. Discrete Online Cross-modal Hashing (DOCH) \cite{zhan2022discrete} has been proposed recently. DOCH not only exploits the similarity between the existing data and the new data, but also considers the fine-grained semantic information to learn the binary codes of the new data. In this paper, we focus on single-modal supervised online hashing tasks.

\section{Fast Online Hashing with Multi-label Projection}
In this section, we first give the notations and the definition of the problem. Then, we describe the details of the related algorithms, including the Neighbor-Preserving Algorithm, Reservoir Sampling Strategy,  Similarity Matrix Construction Algorithm, Hash Functions Updating Algorithm and Optimization Method. Finally, we state the online query process in our model. The whole framework of the proposed model is shown in Fig. \ref{framework} and elaborated in the supplementary document.

\subsection{Problem Definition}
Given a set of images $\mathbf X = [\mathbf{x}_1,...,\mathbf{x}_n]\in\mathbb{R}^{d \times n}$ with corresponding semantic labels $\mathbf L = [l_1,...,l_n]\in\{0,1\}^{c \times n}$, where $n$ denotes the total number of the images, $d$ denotes the dimensionality in the original space and $c$ denotes the total number of the categories. The goal of hashing is to map the image instances to hash codes $\mathbf{B} = [\mathbf{b}_1,...,\mathbf{b}_n]\in\{-1, +1\}^{k\times n}$, where $k$ denotes the hash code length. In order to achieve the goal, we utilize the most common linear projection based hash functions which are defined as
\begin{equation}
\mathbf{B} = sgn(\mathbf{W^T}\mathbf{X}),
\end{equation}
where $\mathbf{W} = [\mathbf{w}_i]_{i=1}^k\in\mathbb{R}^{d\times k}$ denotes the projection matrix and $\mathbf{w}_i$ contributes to the $i$-th hash bit. The sign function is defined as
\begin{equation}
    sgn(x)=
    \left\{
    \begin{array}{ll}
    1, & \mathbf{if}\ x \geqslant 0;\\
    -1, & \mathbf{otherwise}.
    \end{array}
    \right.
\end{equation}
The similarity preserving objective will be defined later.

As for the online learning problem, the data comes in a streaming fashion. Therefore, the examples are not all used at once \cite{lin2019supervised}. For subsequent convenience, we distinguish the related  representations between the new stream data and the accumulated existing data. Specifically, let $\mathbf{X}_s^t = [\mathbf{x}_{s1}^t,...,\mathbf{x}_{sn_t}^t]\in\mathbb{R}^{d \times n_t}$ denote the new input stream data at stage $t$, where $n_t$ denotes the batch size (size of the stream data). The corresponding labels are denoted as $\mathbf{L}_s^t = [l_{s1}^t,...,l_{sn_t}^t]\in\mathbb{R}^{c \times n_t}$. The existing data is denoted as $\mathbf{X}_e^t = [\mathbf{X}_{s}^1,...,\mathbf{X}_{s}^{t-1}] = [\mathbf{x}_{e1}^t,...,\mathbf{x}_{em_t}^t]\in\mathbb{R}^{d \times m_t}$, where $m_t$ denotes the total number of the existing data and $m_t = n_1+...+n_{t-1}$. The corresponding label matrix is denoted as $\mathbf{L}_e^t = [l_{e1}^t,...,l_{em_t}^t]\in\mathbb{R}^{c \times m_t}$. Correspondingly, we denote $\mathbf{B}_s^t = sgn(\mathbf{W}^{t^T}\mathbf{X}_s^t) =  [\mathbf{b}_{s1}^t,...,\mathbf{b}_{sn_t}^t]\in\mathbb{R}^{k \times n_t}$, $\mathbf{B}_e^t = sgn(\mathbf{W}^{t^T}\mathbf{X}_e^t) =  [\mathbf{b}_{e1}^t,...,\mathbf{b}_{em_t}^t]\in\mathbb{R}^{k \times m_t}$ as the learnt binary codes at stage $t$ of the stream data and the existing data, respectively.

\subsection{Neighbor-Preserving Algorithm}
The aim of the neighbor-preserving algorithm is to keep the nearest neighbors of each central point
up to date in consideration of the latest hash functions $\mathbf{W}^t$ at stage $t$. First, the hash codes of the new stream data ($\mathbf{B}_s^t \in \{-1,+1\}^{k\times n_t}$) are computed as
\begin{equation}
\mathbf{B}_s^t = sgn(\mathbf{W}^t\mathbf{^T}\mathbf{X}_s^t).
\end{equation}
Let $\mathbf{X}_C \in \mathbb{R}^{d \times u}$ denote the set of the central points, where $u$ denotes the total number of the central points. Let $\mathbf{X}_{N_i} \in \mathbb{R}^{d \times v}$ denote the nearest neighbors of the $i$-th central point, where $v$ denotes the number of the nearest neighbors of each central point. Hence, $\mathbf{X}_N = [\mathbf{X}_{N_1},...,\mathbf{X}_{N_C}] \in \mathbb{R}^{d \times uv}$ denotes the set of all the potential neighbors. Notice that $\mathbf{X}_N$ is not disjoint. Then, the hash codes of the central points ($\mathbf{B}_C^t$) and the potential points ($\mathbf{B}_N^t$) are calculated as
\begin{equation}
\mathbf{B}_{C}^{t} = sgn(\mathbf{W}^{t^\mathbf{T}}\mathbf{X}_{C}),
\end{equation}
\begin{equation}
\mathbf{B}_N^t = sgn(\mathbf{W}^{t^\mathbf{T}}\mathbf{X}_N).
\end{equation}
For subsequent convenience, we construct a function  which is defined as following:
\begin{equation}
SH(\mathbf{A},\mathbf{B},\alpha) = sort(Hamm(\mathbf{A,B}),\alpha).
\label{equ_SH}
\end{equation}
Suppose $\mathbf{A}\in\mathbb\{0,1\}^{k\times n_a}$ and $\mathbf{B}\in\mathbf\{0,1\}^{k\times n_b}$ denote two hash matrices, where $n_a$ and $n_b$ denote the numbers of hash codes in $\mathbf{A}$ and $\mathbf{B}$, respectively. $Hamm(\mathbf{A,B})\in \mathbb{R}^{n_b\times n_a}$ denotes the Hamming distance matrix. Specifically, the $j$-th column in $Hamm(\mathbf{A,B})$ denotes Hamming distances between $j$-th column in $\mathbf{A}$ and all the columns in $\mathbf{B}$. $sort(Hamm(\mathbf{A,B}),\alpha)\in \mathbb{R}^{\alpha\times n_a}$ returns the \textbf{indices} of the first $\alpha$ neighbors with smallest values in $\mathbf{B}$. Finally, the nearest neighbors of the $i$-th central point ($\mathbf{A}_{N_i}$) are updated as following:
\begin{equation}
\mathbf{X}_{N_i} = \mathbf{X}_ {SH(\mathbf{B}_{C_i}^t,[\mathbf{B}_s^t,\mathbf{B}_{N_i}^t],v)},\ i = 1,2,...,u.
\end{equation}
In this way, the potential neighbors can be updated dynamically with the increasing stream data.

\subsection{Reservoir Sampling Strategy}
After the stream data are accumulated for several rounds, we need to update a part of the central points in the query pool and the corresponding nearest neighbors based on the reservoir sampling strategy. The nearest neighbors of the $i$-th updated central point ($\mathbf{X}_{NU_i}$) are calculated as
\begin{equation}
\mathbf{X}_{NU_i} = \mathbf{X}_ {SH(\mathbf{B}_{CU_i}^t,[\mathbf{B}_e^t,\mathbf{B}_s^t],v)},\ i = 1,...,r,
\end{equation}
where $\mathbf{B}_{CU_i}^t = sgn(\mathbf{W}^{t^\mathbf{T}}\mathbf{X}_{CU_i})$ denotes the hash code of the $i$-th updated central point and $r$ denotes the number of the updated central points.

\subsection{Similarity Matrix Construction Algorithm}

As for multi-label supervision information, traditional supervised methods set the similarity between two examples as one if they share at least one common label, otherwise as zero. Obviously, this kind of binarization on the similarities ignores the detailed label sharing information among the instances. Here, we propose a similarity matrix construction algorithm to solve the above problem. Suppose $[x_i,x_j]$ denotes two examples and $[l(x_i),l(x_j)]$ denotes the corresponding labels. We define $||l(x_i)||$ and $||l(x_j)||$ denote the numbers of the labels that the two instances contain, respectively. Then, we have
\begin{equation}
\mathbf{s}^+(x_i,x_j) = \frac{||l(x_i)\cap l(x_j)||}{||l(x_i)||},
\end{equation}
\begin{equation}
\mathbf{s}^-(x_i,x_j) = \frac{||l(x_i)\cap l(x_j)||}{||l(x_j)||},
\end{equation}
where $||l(x_i)\cap l(x_j)||$ denotes the number of the common labels that $x_i$ and $x_j$ share. Hence, the similarity between $x_i$ and $x_j$ is computed as
\begin{equation}
    \mathbf{s}(x_i,x_j) = \frac{\mathbf{s}^+(x_i,x_j)+\mathbf{s}^-(x_i,x_j)}{2}.
\end{equation}
We further clarify the meanings of these notations. $\mathbf{s}^+(x_i,x_j)$ denotes the similarity between $x_i$ and $x_j$ from $i$-th example's point of view and $\mathbf{s}^-(x_i,x_j)$ denotes the similarity between $x_i$ and $x_j$ from $j$-th example's point of view. Finally, we just take the average of them to define the similarity between the two examples, which means that every instance is treated equally in our algorithm. We also provide an example based on the proposed similarity matrix construction algorithm in the supplementary document.

\subsection{Hash Functions Updating Algorithm}
In this part, we construct an integrated loss function to update the hash functions dynamically. In order to preserve similarity between the original space and Hamming space, we take both the similarity matrix and label information into consideration.

In view of the similarity matrix, the hash functions should be updated by minimizing the error between the similarity matrix and inner product of the hash codes. We utilize Frobenius-norm to express the formulation as
\begin{equation}
\mathop{\min}_{\mathbf{B}_s^t,\mathbf{B}_e^t} \left \| \mathbf{B}_s^{t^\mathbf{T}}\mathbf{B}_e^t - k\mathbf{S^t} \right \|_F^2,
\label{loss_similarity}
\end{equation}
where $\mathbf{S^t}\in\mathbb{R}^{n_t \times m_t}$ is created based on the new stream data $\mathbf{X}_s^t$ and the existing data $\mathbf{X}_e^t$ with the proposed similarity matrix construction algorithm. It is worth noticing that in the single-label case, the similarity is defined as
\begin{equation}
    \mathbf{S}_{ij}
    \left\{
    \begin{array}{ll}
    1, & if\ l(\mathbf{x}_{s_i}^t) = l(\mathbf{x}_{e_j}^t);\\
    -1, & \mathbf{otherwise}.
    \end{array}
    \right.
\end{equation}
Notice that the size of the stream data and existing data may not be the same, which contributes to an asymmetric similarity graph. Furthermore, most of the data pairs are probably dissimilar, which results in a sparse similarity matrix. To solve the above imbalance problem, we add two balance factors $\eta_s$ and $\eta_d$ as weights for similar and dissimilar examples, respectively. We set $\eta_s>>\eta_d$ ($\eta_s = 1.2$, $\eta_d = 0.2$), the Hamming distances among similar pairs are minified, whereas that among dissimilar pairs are enlarged \cite{lin2019towards}.
In addition, we add the quantization error on the new stream data:
\begin{equation}
\left \| \mathbf{W}^t{^\mathbf{T}}\mathbf{X}_s^t - \mathbf{B}_s^t \right \|_F^2.
\label{loss_quantization}
\end{equation}

As for the label information, we construct a label projection loss to make full use of the labels. Specifically, the labels of both the stream data and existing data are projected to approach their corresponding hash codes, which is formulated as
\begin{equation}
\mathop{\min}_{\mathbf{B}_s^t,\mathbf{B}_e^t} \left \| \mathbf{B}_s^{t}-\mathbf{P}^t\mathbf{L}_s^t\right \|_F^2+\left \| \mathbf{B}_e^{t}-\mathbf{P}^t\mathbf{L}_e^t\right \|_F^2, \label{loss_label}
\end{equation}
where $\mathbf{P}^t\in\mathbb{R}^{k \times c}$ denotes the projection matrix to map the labels into hash codes. The label projection loss can make the binary codes more distinguishable and alleviate the imbalance problem caused by the coarse-grained similarity matrices effectively.

By combining Eq. \ref{loss_similarity}, Eq. \ref{loss_quantization} and Eq. \ref{loss_label}, we obtain the overall objective function as
\begin{equation}
\begin{split}
\mathop{\min}_{\mathbf{B}_s^t,\mathbf{B}_e^t,\mathbf{W}^t,\mathbf{P}^t} \left \| \mathbf{B}_s^{t^T}\mathbf{B}_e^t - k\mathbf{S^t} \right \|_F^2 + \sigma\left \| \mathbf{W}^{t^\mathbf{T}}\mathbf{X}_s^t - \mathbf{B}_s^t \right \|_F^2\\
+\theta\left \| \mathbf{B}_s^{t}-\mathbf{P}^t\mathbf{L}_s^t\right \|_F^2+\mu\left \| \mathbf{B}_e^{t}-\mathbf{P}^t\mathbf{L}_e^t\right \|_F^2\\
+\lambda\left \| \mathbf{W}^t\right \|_F^2+\tau\left \| \mathbf{P}^t\right \|_F^2\\
s.t.\ \mathbf{B}_s^t\in\{-1,+1\}^{k \times n_t},\mathbf{B}_e^t\in\{-1,+1\}^{k \times m_t}.
\end{split}
\label{loss}
\end{equation}
where $\sigma,\theta,\mu,\lambda,\tau$ are the parameters to balance the trade-offs among the five learning parts.

\subsection{Optimization Method}
Due to the binary constraints, the optimization problem of Eq. \ref{loss} is non-convex with respect to $\mathbf{B}_s^t,\mathbf{B}_e^t,\mathbf{W}^t,\mathbf{P}^t$. In order to find a feasible solution, we adopt an alternating optimization approach by updating one variable with the rest fixed until convergence.

\textcircled{1} $\mathbf{W}^t\mathbf{-step}:$ By fixing other variables except for  $\mathbf{W}^t$, we update $\mathbf{W}^t$ with a close-formed solution as
\begin{equation}
\mathbf{W}^t=\sigma(\sigma(\mathbf{X}_s^t\mathbf{X}_s^{t^\mathbf{T}}+\lambda\mathbf{I}_d)^{-1}\mathbf{X}_s^t\mathbf{B}_s^{t^\mathbf{T}}),
\end{equation}
where $\mathbf{I}_d\in\mathbb{R}^{d \times d}$ is an identity matrix.

\textcircled{2} $\mathbf{P}^t\mathbf{-step}:$ By fixing other variables except for $\mathbf{P^t}$, we get a closed-form solution of $\mathbf{P}^t$:
\begin{equation}
\mathbf{P}^t=(\mu\mathbf{B}_e^t\mathbf{L}_e^{t^\mathbf{T}} + \theta\mathbf{B}_s^{t}\mathbf{L}_s^{t^\mathbf{T}})(\theta\mathbf{L}_s^{t}\mathbf{L}_s^{t^\mathbf{T}} + \mu\mathbf{L}_e^{t}\mathbf{L}_e^{t^\mathbf{T}} + \tau\mathbf{I}_c)^{-1},
\end{equation}
where $\mathbf{I}_c\in\mathbb{R}^{c \times c}$ is an identity matrix.

\textcircled{3} $\mathbf{B}_e^t\mathbf{-step}:$ By fixing other variables except for  $\mathbf{B}_e^t$, we rewrite Eq.\ref{loss} to obtain the solution of $\mathbf{B}_e^t$ via the following optimization problem
\begin{equation}
\begin{split}
\min_{\mathbf{B}_e^t} \left \| \mathbf{B}_s^{t^\mathbf{T}}\mathbf{B}_e^t \right \|_F^2 + \underbrace{\left \| k\mathbf{S^t} \right \|_F^2}_{const} - 2tr(k\mathbf{B}_e^{t^\mathbf{T}}\mathbf{B}_s^t\mathbf{S^t}) \\
 + \mu(\underbrace{\left \| \mathbf{P}^t\mathbf{L}_e^t \right \|_F^2 + \left \| \mathbf{B}_e^{t}\right \|_F^2}_{const} - 2tr(\mathbf{B}_e^{t^\mathbf{T}}\mathbf{P}^{t}\mathbf{L}_e^{t})) \\
s.t.\ \mathbf{B}_e^t\in\{-1,+1\}^{k \times m_t}.
\end{split}
\end{equation}
where the const items imply that they are not related to solving $\mathbf{B}_e^t$. Inspired by the recent advance on binary codes optimization \cite{xu2020relaxed}, we obtain
\begin{equation}
\begin{split}
\min_{\mathbf{B}_e^t} tr(\mathbf{B}_e^{t^\mathbf{T}}\mathbf{B}_s^t\mathbf{B}_s^{t^\mathbf{T}}\mathbf{B}_e^t)-2tr(\mathbf{B}_e^{t^\mathbf{T}}\mathbf{Z}) \\
=\min_{\mathbf{B}_e^t} tr(\mathbf{B}_e^{t^\mathbf{T}}(\mathbf{B}_s^t\mathbf{B}_s^{t^\mathbf{T}}\mathbf{B}_e^t-2\mathbf{Z})) \\
s.t.\ \mathbf{B}_e^t\in\{-1,+1\}^{k \times m_t}.
\end{split}
\end{equation}
where $\mathbf{Z}=k\mathbf{B}_s^t\mathbf{S}^t-\mu\mathbf{P}^t\mathbf{L}_e^t$. Then, we have the closed form solution of $\mathbf{B}_e^t$ as:
\begin{equation}
\mathbf{B}_e^{t+1}=sgn(2\mathbf{Z}-\mathbf{B}_s^t\mathbf{B}_s^{t^\mathbf{T}}\mathbf{B}_e^t).
\end{equation}

\textcircled{4} $\mathbf{B}_s^t\mathbf{-step}:$ By fixing other variables except for $\mathbf{B}_s^t$, the sub-optimization of Eq.\ref{loss} is equivalent to
\begin{equation}
\begin{split}\label{10}
\min_{\mathbf{B}_s^t} \left \| \mathbf{B}_s^{t^\mathbf{T}}\mathbf{B}_e^t - k\mathbf{S^t} \right \|_F^2+\sigma\left \| \mathbf{W}^{t^\mathbf{T}}\mathbf{X}_s^t - \mathbf{B}_s^t \right \|_F^2 \\
+\theta\left \| \mathbf{B}_s^{t}-\mathbf{P}^t\mathbf{L}_s^t\right \|_F^2 \\
s.t.\ \mathbf{B}_s^t\in\{-1,+1\}^{k \times n_t}.
\end{split}
\end{equation}
The above formulation is equivalent to
\begin{equation}
\begin{split}
\min_{\mathbf{B}_s^t} \left \| \mathbf{B}_s^{t^\mathbf{T}}\mathbf{B}_e^t \right \|_F^2+\underbrace{\left \| k\mathbf{S^t} \right \|_F^2}_{const}-2tr(k\mathbf{S^t}\mathbf{B}_e^{t^\mathbf{T}}\mathbf{B}_s^t) \\
+\sigma(\underbrace{\left \| \mathbf{W}^{t^\mathbf{T}}\mathbf{X}_s^t \right \|_F^2+\left \| \mathbf{B}_s^t) \right \|_F^2}_{const}-2tr(\mathbf{X}_s^{t^\mathbf{T}}\mathbf{W}^t\mathbf{B}_s^t)) \\
+\theta(\underbrace{\left \| \mathbf{P}^t\mathbf{L}_s^t \right \|_F^2+\left \| \mathbf{B}_s^{t}\right \|_F^2}_{const}-2tr(\mathbf{L}_s^{t^\mathbf{T}}\mathbf{P}^{t^\mathbf{T}}\mathbf{B}_s^{t})) \\
s.t.\ \mathbf{B}_s^t\in\{-1,+1\}^{k \times n_t}.
\end{split}
\label{loss_bst}
\end{equation}
Similarly, we ignore irrelevant items to $\mathbf{B}_s^t$. For convenience, the optimization problem in Eq. \ref{loss_bst} is rewritten as
\begin{equation}
\begin{aligned}
&\min_{\mathbf{B}_s^t} \left \| \mathbf{B}_s^{t^\mathbf{T}}\mathbf{B}_e^t \right \|_F^2 - 2tr(\mathbf{G}^\mathbf{T}\mathbf{B}_s^{t}),
\end{aligned}
\label{loss_bst_rewritten}
\end{equation}
where $\mathbf{G}=k\mathbf{B}_e^t\mathbf{S^{t^T}}+\sigma\mathbf{W}^{t^\mathbf{T}}\mathbf{X}_s^t+\theta\mathbf{P}^t\mathbf{L}^t$.
Since it is difficult to optimize $\mathbf{B}_s^t$ directly, we optimize each hash bit one by one. Thus, we obtain a closed form solution for each hash bit by extending  Eq. \ref{loss_bst_rewritten} to the following form:
\begin{equation}
\begin{split}
\min_{\mathbf{\tilde{b}}_s^{t}} \underbrace{\left \| \mathbf{\tilde{b}}_s^{t^\mathbf{T}}\mathbf{\tilde{b}}_e^t \right \|_F^2+\left \| \mathbf{\tilde{B}}_s^{t^T}\mathbf{\tilde{B}}_e^t \right \|_F^2}_{const} + 2tr(\mathbf{\tilde{B}}_s^{t^\mathbf{T}}\mathbf{\tilde{B}}_e^t\mathbf{\tilde{b}}_e^{t^\mathbf{T}}\mathbf{\tilde{b}}_s^t) \\
- 2tr(\mathbf{\tilde{g}}^\mathbf{T}\mathbf{\tilde{b}}_s^{t}) - \underbrace{2tr(\mathbf{\tilde{G}}^\mathbf{T}\mathbf{\tilde{B}}_s^{t})}_{const} \\
=\min_{\mathbf{\tilde{b}}_s^{t}} 2tr((\mathbf{\tilde{B}}_s^{t^\mathbf{T}}\mathbf{\tilde{B}}_e^t\mathbf{\tilde{b}}_e^{t^T}-\mathbf{\tilde{g}}^T)\mathbf{\tilde{b}}_s^{t}),
\end{split}
\end{equation}
where $\mathbf{\tilde{b}}_s^{t}$ and $\mathbf{\tilde{B}}_s^{t}$ denote the hash bits to be updated and fixed, respectively.
This is also suitable for the meanings of  $\mathbf{\tilde{g}}^\mathbf{T}$ and $\mathbf{\tilde{G}}^\mathbf{T}$. Therefore, we obtain a solution for solving $\mathbf{\tilde{b}}_s^{t}$ as
\begin{equation}
\mathbf{\tilde{b}}_s^{t}=sgn(\mathbf{\tilde{g}}-\mathbf{\tilde{b}}_e^{t}\mathbf{\tilde{B}}_e^{t^\mathbf{T}}\mathbf{\tilde{B}}_s^{t}).
\end{equation}

\begin{table*}[t]
\caption{The results on the updating time and query time with 32 and 48 hash bits on two datasets.}\label{tab_query_time}
\vspace{-0.1in}
\centering\label{time}
\tabcolsep=0.3cm

\begin{tabular}{ccccccccc}
\toprule
\multirow{3}{*}{Methods} &\multicolumn{4}{c}{CIFAR-10}    &\multicolumn{4}{c}{FLICKR-25K}  \\ 

		& \multicolumn{2}{c}{Updating time (s)} & \multicolumn{2}{c}{Query time (s)} & \multicolumn{2}{c}{Updating time (s)} & \multicolumn{2}{c}{Query time (s)} \\ 
		
		& 32-bits & 48-bits & 32-bits & 48-bits & 32-bits & 48-bits & 32-bits & 48-bit\\
		\midrule
		OKH & \underline{0.03} & \underline{0.03} & \underline{8.46} & 5.48 & \underline{0.13} & 0.10 & 8.70 & 3.84 \\
		AdaptHash & 0.15 & 0.21 & 8.47 & 5.29 &0.19     &0.11 & 11.3  &4.58 \\
		OSH &0.27    &0.21  &8.79 & \underline{5.27} &0.18   &0.13 & 6.90    &3.10 \\
		MIHash &0.27     &0.25 &\underline{8.46}     &5.29 & 0.19 &0.13   &\underline{6.87} &\underline{3.05} \\
		BSODH &0.26    &0.24 & 8.91    &5.90 & 0.15 &0.10    &7.36 &4.40 \\
		HMOH &0.17    &0.27 &9.25    &5.84 & 0.15 &\underline{0.09}    &7.01  &4.50\\
		FOH & \textbf{0.02}  & \textbf{0.02} & \textbf{2.18}  &\textbf{1.33} & \textbf{0.01} & \textbf{0.01}  & \textbf{2.45} & \textbf{2.85} \\
\bottomrule
\end{tabular}
\vspace{-0.1in}
\end{table*}

To implement the whole algorithm, we first initialize $\mathbf{W}^1$ and $\mathbf{G}^1$ with a standard Gaussian distribution. Then the above four steps are repeated until convergence.

\subsection{Online Query Process}
When a new query $\mathbf{q}\in\mathbb{R}^d$ arrives, suppose the latest hash function is $\mathbf{W}^t$. First, the hash code of the query ($\mathbf{B_q}$) is computed as
\begin{equation}
\mathbf{B_q} =  sgn(\mathbf{W}^{t^\mathbf{T}}\mathbf{q}).
\end{equation}
Given the hash codes of the central points $\mathbf{B}_C^t$, we search for the nearest central points of the query point in the query pool and return the corresponding potential neighbors of $\mathbf{q}$ ($\mathbf{X}_P$) by
\begin{equation}
\mathbf{X}_P = \mathbf{X}_{N_{SH(\mathbf{B_q}^t,\mathbf{B}_C^t,\beta)}},
\end{equation}
where $\beta$ denotes the number of the returned central points. Then, the hash codes of the potential neighbors ($\mathbf{B}_P^t$) are calculated as
\begin{equation}
\mathbf{B}_P^t =  sgn(\mathbf{W}^{t^\mathbf{T}}\mathbf{X}_P^t).
\end{equation}
Finally, the retrieval results ($\mathbf{R}_K\in\mathbb{R}^{d\times K}$) are computed as
\begin{equation}
\mathbf{R}_K = \mathbf{X}_{SH(\mathbf{B_q}^t,\mathbf{B}_P^t,K)},
\end{equation}
where $K$ denotes the number of the required returned nearest neighbors of the query point.

Throughout the whole online query process, we find that the query time is dramatically reduced. One reason is that the number of data whose binary codes need to be updated is  declined. The other reason is that the online Hamming distances calculation time and the hash codes sorting time are both decreased.

\section{Experiments}
In this section, we conduct experiments on two common datasets: CIFAR-10 \cite{krizhevsky2009learning} and FLICKR-25K \cite{huiskes2008mir} to verify the efficiency and effectiveness of the proposed Fast Online Hashing (FOH).

\subsection{Datasets and Evaluation Protocols}
CIFAR-10 is a widely used image retrieval dataset containing 60,000 images in 10 different categories. We randomly select 1,000 examples as the query set and the rest is regarded as the base set. Furthermore, 20,000 images are randomly sampled from the base set as the training set. In order to simulate the stream data, we divide the training set into 10 blocks with 2,000 examples in each block.

FLICKR-25K contains 25,000 images annotated by 24 provided labels. We select the images that own at least 20 tags. Hence, 20,015  examples are obtained for our experiment. We randomly select 2,000 examples as the test and the rest are served as both the training set and base set. For online hashing, we divide the training set into nine blocks, with the first eight blocks each containing 2,000 examples and the ninth block containing 2,015 examples.

We evaluate the accuracy performance of the online hashing methods using three criteria: Recall@$k$, Precision@$k$ and mAP. Recall@$k$ is calculated by the percentage of the true neighbors in the whole true neighbor set, whereas Precision@$k$ is calculated using the percentage of the true neighbors in the result set. The mAP is calculated based on the mean value of the Precision@$k$ for all true neighbors.

\subsection{Baselines and Settings}
In order to reveal the superiority of FOH, we compare with several state-of-the-art online hashing baselines, including Online Kernel Hashing (OKH) \cite{huang2013online}, Adaptive hashing (AdaptHash) \cite{cakir2015adaptive}, Online Supervised Hashing (OSH) \cite{Cakir2015online}, OH with Mutual Information (MIHash) \cite{cakir2017mihash},  Towards Optimal Discrete Online Hashing with Balanced Similarity (BSODH) \cite{lin2019towards} and Hadamard Matrix Guided Online Hashing (HMOH) \cite{lin2020hadamard}. We use a pre-trained VGG16 \cite{vgg16} for all the baselines to extract the original real-value image features.

Here, we provide the exact values of the parameter configurations in Tab. \ref{tab_parameter}. Review that $u$ denotes the number of the central points in the query pool, $v$ denotes the number of the nearest neighbors of each central point, $\beta$ denotes the number of the returned central points when a new query arrives, $\{\sigma, \theta, \mu,  \lambda, \tau\}$ denotes the hyper-parameters in the objective function.

\begin{table}[h!]
\vspace{-0.05in}
\caption{Parameter configurations on two datasets.}
\vspace{-0.1in}
\centering
\tabcolsep=0.12cm
\label{tab_parameter}
\begin{tabular}{ccccccccc}
\toprule
Dataset & u & v & $\beta$ & $\sigma$ & $\theta$ & $\mu$ & $\lambda$ & $\tau$ \\
\midrule
CIRAR-10 & 500 & 500 & 10 & 0.8 & 1.2 & 0.5 & 0.6 & 0.6 \\
FLICKR-25K & 200 & 500 & 10 & 0.8 & 1.5 & 0.5 & 0.5 & 5 \\
\bottomrule
\end{tabular}
\vspace{-0.2in}
\end{table}

\begin{table*}[t]
\caption{The mAP scores of different online hashing methods with hash bits from 16 to 128 bits on two datasets.}\label{tab_map}
\vspace{-0.1in}
\centering
\begin{tabular}{ccccccccccc}
\toprule
\multirow{2}*{Methods} & \multicolumn{5}{c}{CIFAR-10} & \multicolumn{5}{c}{FLICKR-25K} \\
& 16-bits & 32-bits & 48-bits & 64-bits & 128-bits & 16-bits & 32-bits & 48-bits & 64-bits & 128-bits \\
\midrule
OKH & 0.134 & 0.223 & 0.252 & 0.268 & 0.350 & 0.531 & 0.536 & 0.532 & 0.535 & 0.537 \\
AdaptHash & 0.138 & 0.216 & 0.297 & 0.305 & 0.293 & 0.544 & 0.545 & 0.547 & 0.548 & 0.555  \\
OSH & 0.126 & 0.129 & 0.131 & 0.127 & 0.125 & 0.540 & 0.542 & 0.547 & 0.550 & 0.561 \\
MIHash & 0.640 & 0.675 & 0.668 & 0.667 & 0.664 & 0.535 & 0.539 & 0.541 & 0.545 & 0.546 \\
BSODH & 0.604 & 0.689 & 0.656 & 0.709 & 0.711 & 0.535 & 0.540 & 0.542 & 0.547 & 0.55 \\
HMOH & \textbf{0.732} & \underline{0.723} & \underline{0.734} & \underline{0.737} & \underline{0.749} & \underline{0.548} & \underline{0.551} & \underline{0.558} & \underline{0.561} & \underline{0.565} \\
FOH & \underline{0.685} & \textbf{0.734} & \textbf{0.746} & \textbf{0.758} & \textbf{0.763} & \textbf{0.585} & \textbf{0.604} & \textbf{0.605} & \textbf{0.610} & \textbf{0.615} \\
\bottomrule
\end{tabular}
\end{table*}

\begin{figure*}
\vspace{-0.15in}
\begin{center}
\mbox{
\includegraphics[width=2.1in]{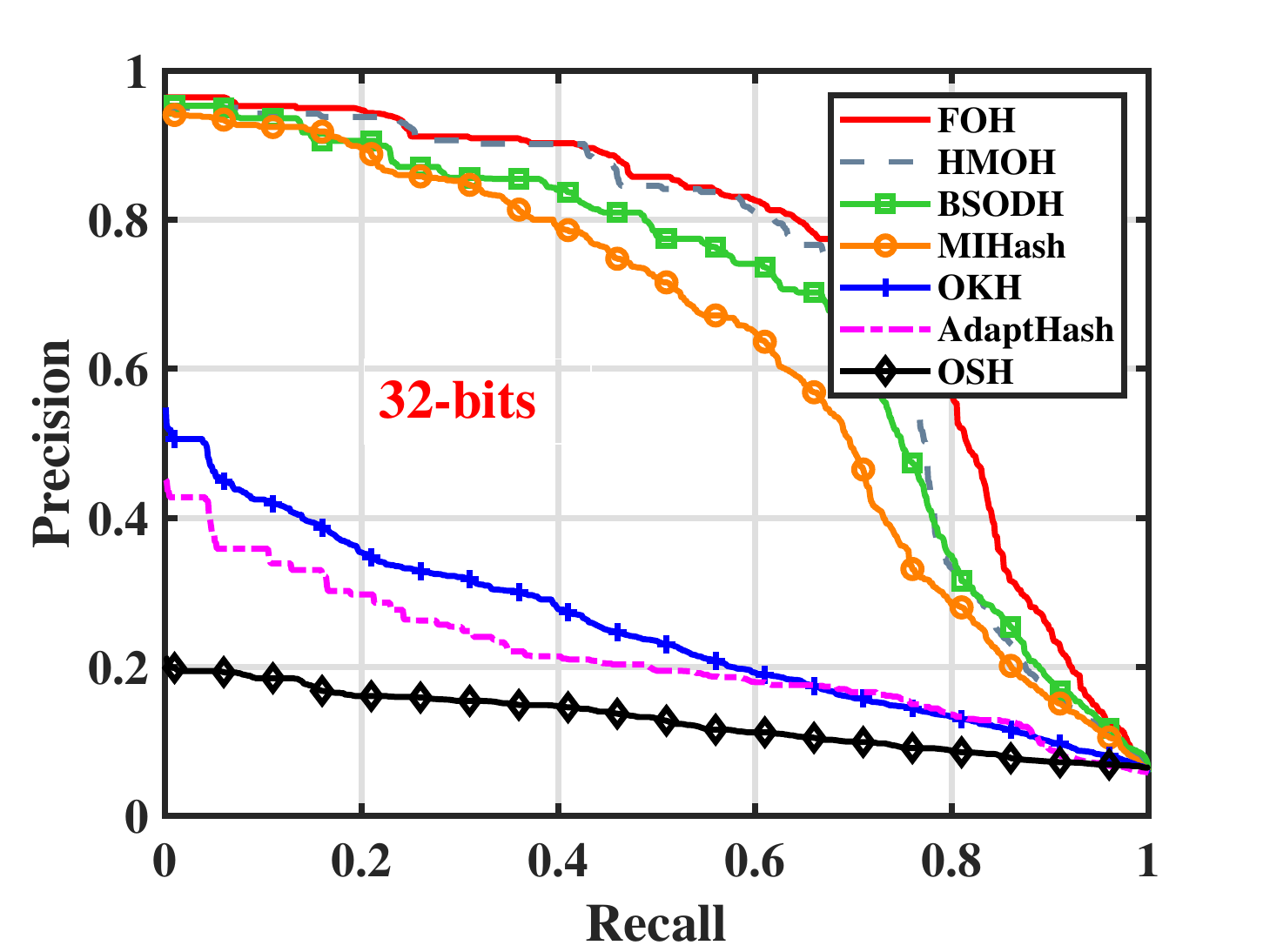}
\includegraphics[width=2.1in]{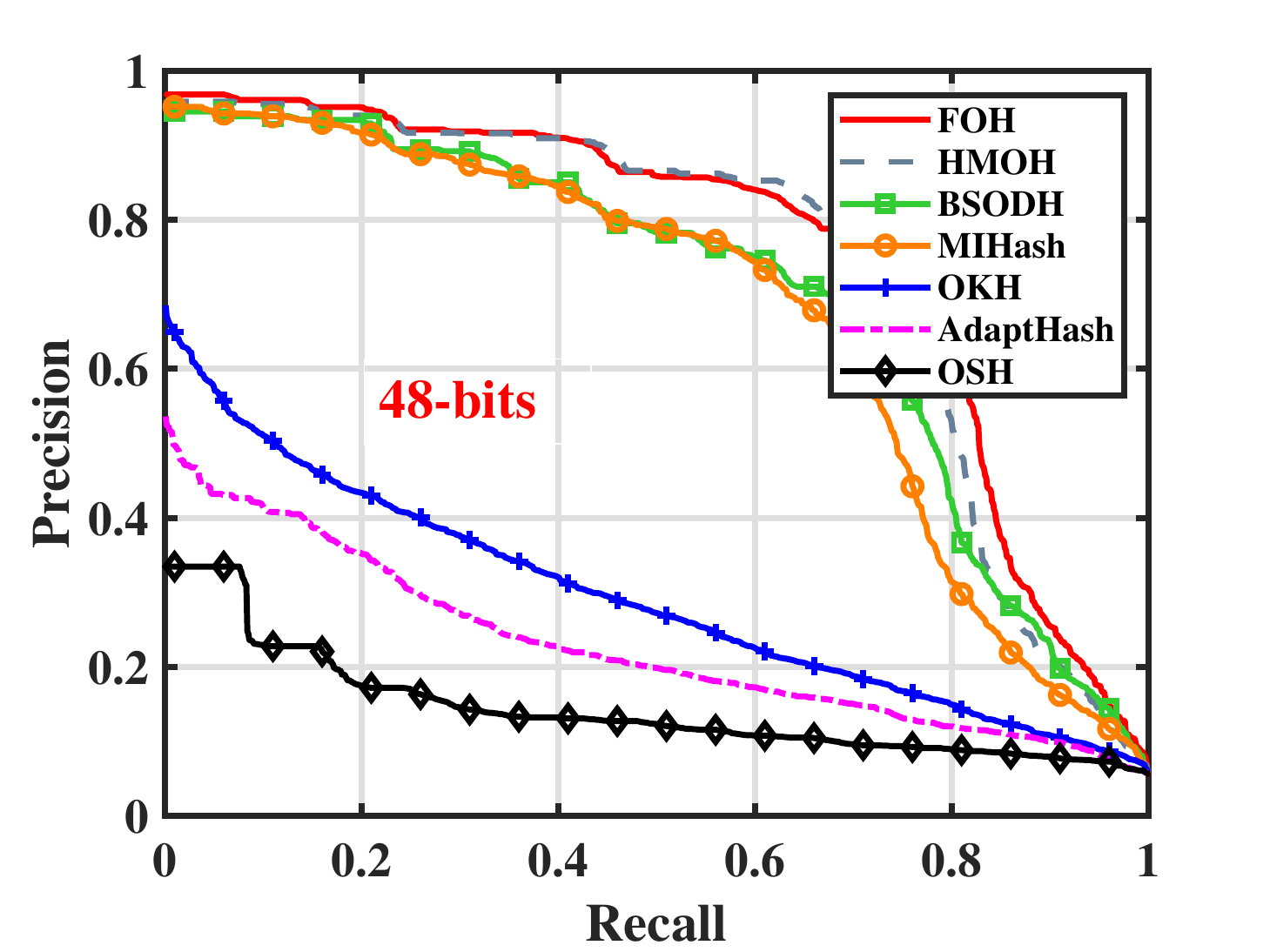}
\includegraphics[width=2.1in]{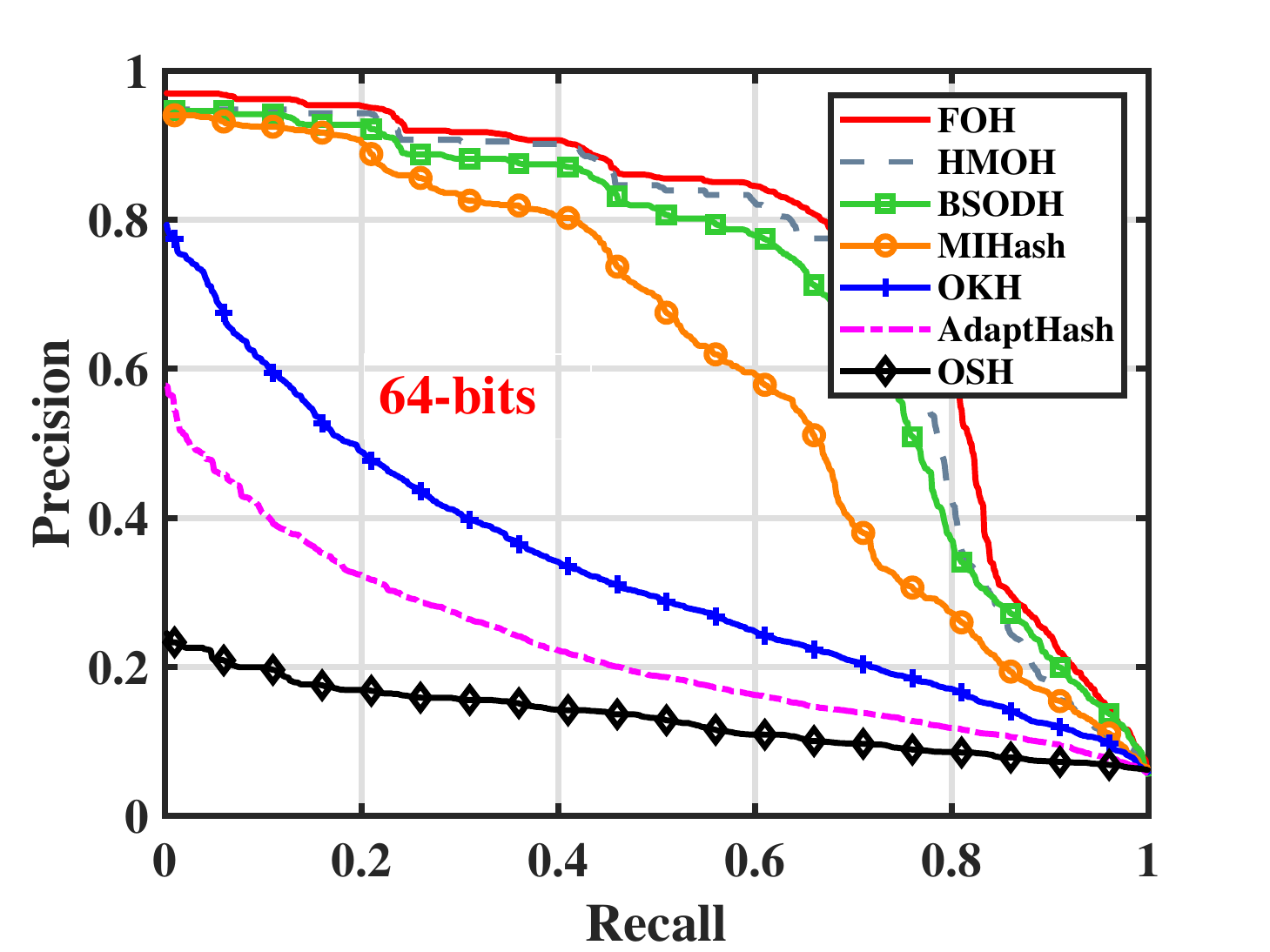}
}
\end{center}
\vspace{-0.1in}
\caption{Precision-recall curves of different online hashing methods with 32, 48 and 64 hash bits on CIFAR-10. }\label{fig_pr}
\vspace{-0.1in}
\end{figure*}

\subsection{Results and Analysis}
To verify the efficiency of FOH, we conduct experiments on the hash table updating time and online query time on two datasets. As shown in Tab. \ref{tab_query_time}, the baselines performs similarly. However, FOH obtains dramatic reduction on both the updating time and query time. Specifically, FOH yields up to 6.28 seconds and 4.42 seconds less query time on CIFAR-10 and FLICKR-25K, respectively. We also compare the training time of FOH with other baselines in Tab. \ref{tab_training_time}. We find that the training time of FOH is relatively low.

In addition, we also make a comparison on the retrieval accuracy of FOH and state-of-the-art baselines. Tab. \ref{tab_map} shows the mAP scores with hash bits from 16 to 128 bits on two datasets. Fig. \ref{fig_pr} shows the precision-recall curves of different online hashing methods with 32, 48 and 64 hash bits on CIFAR-10. It is obvious that FOH gets the highest accuracy with most of the hash bits, which reveals that FOH is competitive in consideration of accuracy. More experimental results and analysis are displayed in the supplementary document.

\begin{table}[t]
\caption{The results on the training time with 32 and 48 hash bits on two datasets.}\label{tab_training_time}
\vspace{-0.1in}
\centering\label{time}
\tabcolsep=0.15cm
\begin{tabular}{ccccc}
\toprule
\multirow{3}{*}{Methods}
&\multicolumn{4}{c}{Training time (s)}\\ 
&\multicolumn{2}{c}{CIFAR-10}    &\multicolumn{2}{c}{FLICKR-25K}  \\ 
		& 32-bits & 48-bits & 32-bits & 48-bits \\
		\midrule
		OKH & 4.78 & 5.62 & 4.70 & 5.30   \\
		AdaptHash & 20.8  & 42.3 & 12.3     &19.5      \\
		OSH &93.5    &128  &60.1    &90.2      \\
		MIHash &120     &152 &80.2     &102     \\
		BSODH &20.6    &21.6 & 3.09    &3.22    \\
		HMOH &7.04    & 10.5 &2.85    &3.09      \\
		FOH & 19.5  &21.7 & 3.40  &4.27      \\
\bottomrule
\end{tabular}
\vspace{-0.2in}
\end{table}

\subsection{Ablation Study}
We configure three variants of FOH to investigate the impacts on accuracy: FOH-Q that removes the query pool, FOH-L that constructs the loss function without label projection, FOH-S that utilizes the traditional similarity matrix in the multi-label case.

\begin{table}[b]
 \centering
 \tabcolsep=0.35cm
 \vspace{-0.05in}
 \caption{The mAP scores of different variants of FOH with 32 and 48 hash bits on two datasets.}\label{tab_ablation}
 \vspace{-0.1in}
 \label{table3}
 \small
 \begin{tabular}{ccccc}
  \toprule
  \multirow{2}{*}{Variants}&\multicolumn{2}{c}{CIFAR-10}&\multicolumn{2}{c}{FLICKR-25K} \\
  & 32-bits & 48-bits & 32-bits & 48-bits  \\
  \midrule
  FOH-Q  & 0.747 & 0.759 & 0.595 & 0.599 \\
  FOH-L & 0.717 & 0.737 & 0.583 & 0.592 \\
  FOH-S  & - & - & 0.554 &0.555 \\
  FOH & 0.734 & 0.746 & 0.598 & 0.605 \\
  \bottomrule

 \end{tabular}
\vspace{-0.1in}
\end{table}

As reported in Tab. \ref{tab_ablation}, we can observe that FOH-L and FOH-S both contribute sufficiently to performance improvement. In comparison with FOH-Q, we find that there is little difference on accuracy by building the query pool. However, it can speed up the query process dramatically. In comparison with FOH-L, we confirm that the label projection loss is crucial for learning. FOH-S shows the biggest performance gap with FOH (4.4\% and 5\% improvement on 32 and 48 hash bits, respectively), which demonstrates that it is indeed effective to gain accuracy by utilizing the proposed similarity matrix construction algorithm.

Tab. \ref{tab_updating} shows how the central points updating frequency affects the retrieval accuracy on two datasets with 32 hash bits (2-b denotes updating the central points when 2 batches of stream data come). Ther are 10 batches in total in CIFAR-10 and 9 batches in total in FLICKR-25K. We find that the mAP drop is very small with the decrease of the updating frequency of the central points. Our imple\-mentation of this paper is publicly available on GitHub at: https://github.com/caoyuan57/FOH.

\begin{table}[t]
 \centering
 \tabcolsep=0.12cm
 \caption{The mAP scores of different updating frequencies of the central points with 32 hash bits on two datasets.}\label{tab_updating}
 \vspace{-0.1in}
 \label{table3}
 \small
 \begin{tabular}{cccccc}
  \toprule
  Datasets & 1-b & 2-b & 3-b & 4-b & 5-b \\
  \midrule
  CIFAR-10 & 0.73403 & 0.72992 & 0.72828 & 0.72626 & 0.72529 \\
  FLICKR-25K & 0.60448 & 0.60134 & 0.59955 & 0.59791 & 0.59135\\
  \bottomrule
 \end{tabular}
\vspace{-0.2in}
\end{table}

\section{Conclusion}
In this paper, we propose a novel fast online hashing model that can speed up the online query time dramatically. To achieve this goal, we build a query pool to retain the potential neighbors with the proposed neighbor-preserving algorithm and reservoir sampling strategy. Furthermore, in order to make full use of the supervision information in the multi-label case, we present a similarity construction algorithm. In the end, an integrated loss function is constructed in consideration of both the similarity matrix and label projection, which contributes to more discriminative hash codes. The experimental results show significant reduction in online query time with competitive retrieval accuracy on two common datasets.

\section{Acknowledgment}
This work is supported by the NSFC Grant Nos. 62202438, 62172090; the Natural Science Foundation of Shandong Province Grant No. ZR2020QF041; the 69th batch of China Postdoctoral Science Foundation Grant No. 862105020017; the 22th batch of ISN Open Fund Grant No. ISN22-21; the CAAI-Huawei MindSpore Open Fund; the Alibaba Group through Alibaba Innovative Research Program; the Fundamental Research Funds for the Central Universities Grant No. 842113037. We thank the Big Data Computing Center of Southeast University for providing the facility support on the numerical calculations in this paper.

\bibliography{bibfile}
\end{document}